\documentclass[conference]{IEEEtran}
\IEEEoverridecommandlockouts
\usepackage{cite}
\usepackage{amsmath,amssymb,amsfonts}
\usepackage{algorithmic}
\usepackage{graphicx}
\graphicspath{{figures/}}
\usepackage{textcomp}
\usepackage{subfigure}
\usepackage{color}
\usepackage{amsmath,amsfonts,amssymb}
\def\BibTeX{{\rm B\kern-.05em{\sc i\kern-.025em b}\kern-.08em
    T\kern-.1667em\lower.7ex\hbox{E}\kern-.125emX}}

\usepackage{listings}
\newcommand{\revised}[1]{\textcolor{black}{#1}}

\usepackage{xcolor}
\usepackage{comment}
\usepackage{wrapfig}
\usepackage{url}

\definecolor{codegreen}{rgb}{0,0.6,0}
\definecolor{codegray}{rgb}{0.5,0.5,0.5}
\definecolor{codepurple}{rgb}{0.58,0,0.82}
\definecolor{backcolour}{rgb}{0.95,0.95,0.92}

\lstdefinestyle{mystyle}{
    backgroundcolor=\color{backcolour},   
    commentstyle=\color{codegreen},
    keywordstyle=\color{magenta},
    numberstyle=\tiny\color{codegray},
    stringstyle=\color{codepurple},
    basicstyle=\ttfamily\footnotesize,
    breakatwhitespace=false,         
    breaklines=true,                 
    captionpos=t,                    
    keepspaces=true,                 
    numbers=left,                    
    numbersep=5pt,                  
    showspaces=false,                
    showstringspaces=false,
    showtabs=false,                  
    tabsize=2
}

\usepackage[switch]{lineno} 

\begin{document}
\pagestyle{plain} 

\author{\IEEEauthorblockN{Romit Maulik}
\IEEEauthorblockA{Argonne National Laboratory\\
rmaulik@anl.gov}
\and
\IEEEauthorblockN{Romain Egele}
\IEEEauthorblockA{École Polytechnique\\ 
romain.egele@polytechnique.edu} 
\and
\IEEEauthorblockN{Bethany Lusch}
\IEEEauthorblockA{Argonne National Laboratory\\
blusch@anl.gov}
\and
\IEEEauthorblockN{Prasanna Balaprakash}
\IEEEauthorblockA{Argonne National Laboratory\\
pbalapra@anl.gov}
}

\title{Recurrent Neural Network Architecture Search for Geophysical Emulation}

\maketitle

\begin{abstract}
Developing surrogate geophysical models from data is a key research topic in atmospheric and oceanic modeling because of the large computational costs associated with numerical simulation methods. Researchers have started applying a wide range of machine learning models, in particular neural networks, to geophysical data for forecasting without these constraints. Constructing neural networks for forecasting such data is nontrivial, however, and often requires trial and error. To address these limitations, we focus on developing proper-orthogonal-decomposition-based long short-term memory networks (POD-LSTMs). We develop a scalable neural architecture search for generating stacked LSTMs to forecast temperature in the NOAA Optimum Interpolation Sea-Surface Temperature data set. Our approach identifies POD-LSTMs that are superior to manually designed variants and baseline time-series prediction methods. We also assess the scalability of different architecture search strategies on up to 512 Intel Knights Landing nodes of the Theta supercomputer at the Argonne Leadership Computing Facility. 
\end{abstract}

\begin{IEEEkeywords}
LSTMs, AutoML, Emulators, Geophysics
\end{IEEEkeywords}

\section{Introduction}

Geophysical forecasting remains an active area of research because of its profound implications for economic planning, disaster management, and adaptation to climate change. Traditionally, forecasts for geophysical applications have relied on the confluence of experimental observations, statistical analyses, and high-performance computing. However, the high-performance computing aspect of forecasting has traditionally been limited to ensemble partial differential  equation (PDE)-based forecasts of different weather models (for example, \cite{skamarock2008description,rogers2009ncep,yoo2013diagnosis}). Recently, there has been an abundance of publicly available weather data  through modern techniques for remote sensing, experimental observations, and data assimilation within numerical weather predictions. Consequently, many researchers have attempted to utilize data for more effective forecasts of geophysical processes \cite{schneider2017earth,gentine2018could,brenowitz2018prognostic,rasp2018deep,o2018using}. For example, researchers have started building data-driven forecasts by emulating the evolution of the weather \emph{nonintrusively} \cite{chattopadhyay2020predicting,chattopadhyay2019data,chattopadhyay2019analog}. In this approach, forecasts are based on data-driven models by eschewing numerical equations. One rationale for these types of predictions is  that equation-based forecasts are inherently limited since they do not capture all the relevant physical processes of the atmosphere or the oceans. More important, the data-driven models are attractive because they promise the possibility of overcoming the traditional limitations of equation-based models based on numerical stability and time to solution. Indeed, nonintrusive surrogate models for PDE-based systems have found popularity in many engineering applications \cite{taira2017modal,wang2019non,maulik2020reduced,qian2020lift} because they have been successful in reducing computational simulation campaigns for product design or in complex systems control
\cite{proctor2016dynamic,peitz2019multiobjective,noack2011reduced,rowley2017model}. 

\begin{figure*}[!ht]
    \centering
    \includegraphics[width=0.75\textwidth]{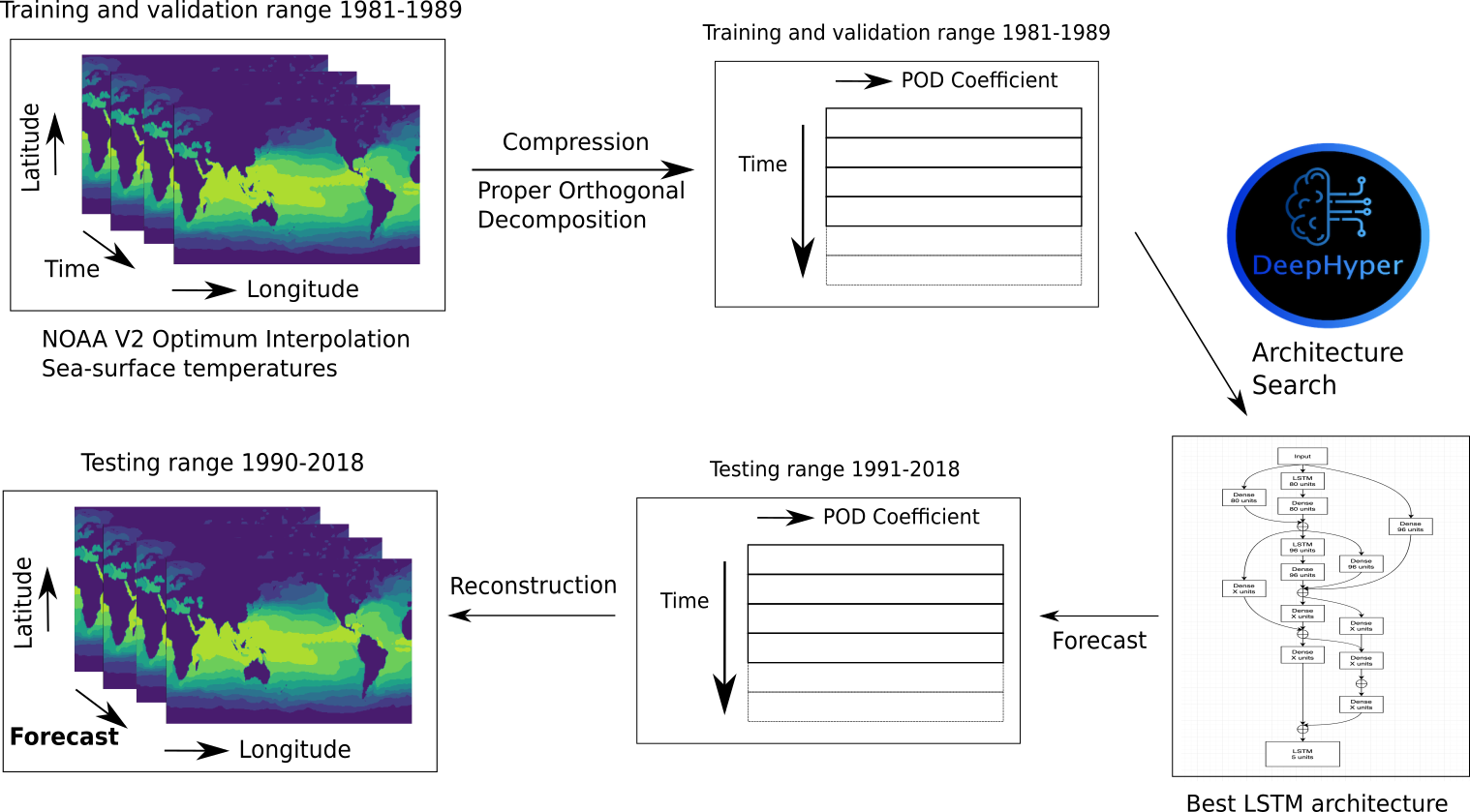}
    \caption{Our proposed NAS approach for  automated POD-LSTM development. Snapshots of spatiotemporally varying training data are compressed by using proper orthogonal decomposition to generate reduced representations that vary with time. These representations (or coefficients) are used to train stacked LSTMs that can forecast on test data. The POD basis vectors obtained from the training data are retained for  reconstruction using the forecast coefficients.
    }
    \label{fig:overall_schematic}
\end{figure*}

We focus on a particularly promising approach for nonintrusive modeling (or forecasting) involving the use of linear dimensionality reduction followed by recurrent neural network time evolution \cite{mohan2018deep,pawar2019deep}. This forecast technique compresses the spatiotemporal field into its dominant principal components by using  proper orthogonal decomposition (POD) (also known as principal components analysis) \cite{kosambi2016statistics,berkooz1993proper}. Following this, the coefficients of each component are evolved by using a time series method. In recent literature, long short-term memory networks (LSTMs), a variant of recurrent neural networks, have been used extensively for  modeling  temporally varying POD coefficients \cite{rahman2019nonintrusive,maulik2020time}. The construction of an LSTM architecture for this purpose is generally based on trial and error, requires human expertise, and consumes significant development time. To address these limitations, we devise an automated way of developing POD-LSTMs using a neural architecture search (NAS) for a real-world geophysical data set, the NOAA Optimum Interpolation Sea-Surface Temperature (SST) data set, which represents a contribution over past POD-LSTM studies that have studied academic data sets alone. In particular, we leverage the NAS framework of DeepHyper \cite{balaprakash2019scalable} to automate the discovery of stacked LSTM architectures that evolve in time POD coefficients  for spatiotemporal data sets. DeepHyper is a scalable open-source hyperparameter and NAS package that was previously assessed for automated discovery of fully connected neural networks on tabular data. In this study, we extend DeepHyper's capabilities for discovering stacked LSTM architectures by parameterizing the space of stacked LSTM architectures as a directed acyclic graph. We adopt the scalable infrastructure of DeepHyper using different search methods for POD-LSTM development. A schematic that describes our overall approach is shown in Figure \ref{fig:overall_schematic}. The main contributions of this work are  as follows.

\begin{itemize}
    \item We develop an automated NAS approach to generate stacked LSTM architectures for POD-LSTM to forecast the global sea-surface temperature on the NOAA Optimum Interpolation SST data set.
    \item We improve the scalability of the NAS approach within DeepHyper by implementing aging evolution, an asynchronous evolutionary algorithm; and we demonstrate its efficacy for developing POD-LSTM. 
    \item We compare aging evolution with reinforcement learning and random search methods at scale on up to 512 nodes of the Theta supercomputer and show that the proposed approach has \revised{better scaling and node utilization}.
    \item We show that automatically obtained POD-LSTMs compare favorably with manually designed variants and baseline machine learning forecast tools. 
\end{itemize}

\section{Data set and preprocessing}

In this section we describe the data set and the POD technique we used for data compression.

\subsection{NOAA Optimum Interpolation Sea-Surface Temperature data set}

For our geophysical emulation we utilize the open-source NOAA Optimum Interpolation SST V2 data set.\footnote{Available at https://www.esrl.noaa.gov/psd/}  Seasonal fluctuations in this data set cause strong periodic structure, although complex ocean dynamics still lead to rich phenomena. Temperature snapshot data is available on a weekly basis on a one-degree grid. This data set has previously been used in data-driven forecasting and analysis tasks (for instance, see \cite{kutz2016multiresolution,callaham2019robust}) particularly from the point of view of identifying seasonal and long-term trends for ocean temperatures by latitude and longitude. Each ``snapshot'' of data comprises  a temperature field in an array of size 360 $\times$ 180 (i.e., the longitudes and latitudes of a one-degree resolution grid), which corresponds to the average sea-surface temperature magnitude for that week. Prior to its utilization for forecasting, a mask is used to remove missing locations in the array that correspond to the land area. The nonzero data points then are flattened to obtain an $\mathbb{R}^Z$-dimensional vector as our final snapshot for a week.

This data is available from October 22, 1981, to June 30, 2018 (i.e., 1,914 snapshots). We utilize the time period of October 22, 1981, to December 31, 1989, for training and validation (427 snapshots). The remaining data set (i.e, 1990 to 2018) is used for testing (1,487 snapshots). Note that this breakdown of the data set into training and testing is commonly used \cite{callaham2019robust} and the 8-year training period captures seasonal as well as subdecade trends in the data set. The training data is utilized to obtain data points given by a window of inputs and a window of outputs corresponding to the desired task of forecasting the future, given observations of the past sea-surface temperatures. Further specifics of the forecasting (i.e., the window of history interpreted as input and the length of the forecast) will be discussed in Section \ref{comp_fore}. These data points are then split into training and validation sets. We note that this forecast is performed non-autoregressively---that is,  the data-driven method \emph{is not} utilized for predictions beyond the desired window size. Since this data set is produced by combining local and satellite temperature observations, it represents an attractive forecasting task for data-driven methods.

\subsection{Compression and forecasting}
\label{comp_fore}

Here, we first review the POD technique for the construction of a reduced basis \cite{kosambi2016statistics,berkooz1993proper} for data compression (see \cite{taira2017modal} for POD and its relationship with other dimension-reduction techniques). The POD procedure involves identifying a space that approximates snapshots of a signal optimally with respect to the $L^2-$norm. The process of orthogonal basis ($\boldsymbol{\vartheta}$) generation commences with the collection of snapshots in the \emph{snapshot matrix},

\begin{align}
\mathbf{S} = [\begin{array}{c|c|c|c}{\hat{\mathbf{q}}^{1}_h} & {\hat{\mathbf{q}}^{2}_h} & {\cdots} & {\hat{\mathbf{q}}^{N_{s}}_h}\end{array}] \in \mathbb{R}^{N_{h} \times N_{s}},
\end{align}
where $N_s$ is the number of snapshots and $\hat{\mathbf{q}}^i_h \in \mathbb{R}^{N_h}$ corresponds to an individual $\mathbb{R}^{N_h}$ degree-of-freedom snapshot in time of the discrete solution domain with the mean value removed, namely,

\begin{align}
\begin{gathered}
\hat{\mathbf{q}}^i_h = \mathbf{q}^i_h - \mathbf{\bar{q}}_h, \quad
\mathbf{\bar{q}}_h = \frac{1}{N_s} \sum_{i=1}^{N_s} \mathbf{q}^i_h,
\end{gathered}
\end{align}
where $\overline{\mathbf{q}}_h \in \mathbb{R}^{N_h}$ is the time-averaged solution field. We note that $\mathbf{q}^i_h$ may be assumed to be any multidimensional signal that is subsequently flattened. Within the context of our geophysical data sets, these correspond to the flattened land or sea-surface temperature snapshots. Our POD bases can then be extracted efficiently through the method of snapshots where we solve the eigenvalue problem on the correlation matrix $\mathbf{C} = \mathbf{S}^T \mathbf{S} \in \mathbb{R}^{N_s \times N_s}$. Then

\begin{align}
\begin{gathered}
\mathbf{C} \mathbf{W} = \mathbf{W} \Lambda,
\end{gathered}
\end{align}
where $\Lambda = \operatorname{diag}\left\{\lambda_{1}, \lambda_{2}, \cdots, \lambda_{N_{s}}\right\} \in \mathbb{R}^{N_{s} \times N_{s}}$ is the diagonal matrix of eigenvalues and $\mathbf{W} \in \mathbb{R}^{N_{s} \times N_{s}}$ is the eigenvector matrix. Our POD basis matrix can then be obtained by
\begin{align}
\begin{gathered}
\boldsymbol{\vartheta} = \mathbf{S} \mathbf{W} \in \mathbb{R}^{N_h \times N_s}.
\end{gathered}
\end{align}
In practice, a reduced basis $\boldsymbol{\psi} \in \mathbb{R}^{N_h \times N_r}$ is built by choosing the first $N_r$ columns of $\boldsymbol{\vartheta}$ for the purpose of efficient reduced-order models, where $N_r \ll N_s$. This reduced basis spans a space given by
\begin{align}
\mathbf{X}^{r}=\operatorname{span}\left\{\boldsymbol{\psi}^{1}, \dots, \boldsymbol{\psi}^{N_r}\right\}.
\end{align}
The coefficients of this reduced basis (which capture the underlying temporal effects) may be extracted as

\begin{align}
\begin{gathered}
\mathbf{A} = \boldsymbol{\psi}^{T} \mathbf{S} \in \mathbb{R}^{N_r \times N_s}.
\end{gathered}
\end{align}
The POD approximation of our solution is then obtained via

\begin{align}
\hat{\mathbf{S}} =  [\begin{array}{c|c|c|c}{\tilde{\mathbf{q}}^{1}_h} & {\tilde{\mathbf{q}}^{2}_h} & {\cdots} & {\tilde{\mathbf{q}}^{N_{s}}_h}\end{array}] \approx \boldsymbol{\psi} \mathbf{A} \in \mathbb{R}^{N_h \times N_s},
\end{align}
where $\tilde{\mathbf{q}}_h^i \in \mathbb{R}^{N_h}$ corresponds to the POD approximation to $\hat{\mathbf{q}}_h^i$. The optimal nature of reconstruction may be understood by defining the relative projection error,

\begin{align}
\frac{\sum_{i=1}^{N_{s}}\left\|\hat{\mathbf{q}}^i_h-\tilde{\mathbf{q}}^i_h \right\|_{\mathbb{R}^{N_{h}}}^{2}}{\sum_{i=1}^{N_{s}}\left\|\hat{\mathbf{q}}^i_h\right\|_{\mathbb{R}^{N_{h}}}^{2}}=\frac{\sum_{i=N_r+1}^{N_{s}} \lambda_{i}^{2}}{\sum_{i=1}^{N_{s}} \lambda_{i}^{2}},
\end{align}
which shows that with increasing retention of POD bases, increasing reconstruction accuracy may be obtained.

Following compression, the overall forecast task may be formulated after recognizing that the rows of $\mathbf{A}$ contain information about the different POD modes and the columns correspond to their varying information in time. Therefore, one approach for forecasting is to predict the evolution of the $N_r$ coefficients in time. Once a forecast has been performed, the first $N_r$ bases may be used to reconstruct the snapshot (in the future). A popular approach for this forecasting task is to use data-driven time series methods. This is motivated by the fact that equations for the evolution of coefficients are nontrivial in the reduced basis. We can then generate training data by extracting the coefficients in $\mathbf{A}$ in a windowed-input and windowed-output form to completely define our forecast task. The state at a given time is represented by an $N_r$-dimensional column vector of POD coefficients. We fix the value of $N_r=5$, which captures approximately 92 \% of the variance of the data. While a larger value of $N_r$ would lead to improved reconstruction of small-scale features, training a stable data-driven model to predict the lower-energy modes requires additional treatment. 
As we demonstrate in Sec. \ref{science-results}, however, setting $N_r=5$ is sufficient to capture the seasonal and long-term trends in sea-surface temperature. Given $N_s$ snapshots of data, we choose every subinterval of width $2K$ as an example, where $K$ snapshots are the input and $K$ snapshots are the output. We utilize a randomly sampled 80\% of examples for training and utilize the remaining 20\% for validation. For our NOAA SST data set, we have $N_s=427$ snapshots, and we choose $K=8$, resulting in 1,111 examples. We note that we avoid any potential data leakage by testing on data points generated in entirely different years of our data (no overlap between training and testing data).

\section{NAS using DeepHyper}

In this section, we describe the stacked LSTM search space for POD-LSTM and the NAS methods employed to explore it. 

\subsection{Stacked LSTM search space}

\begin{figure}
    \centering
    \includegraphics[width=0.36\textwidth]{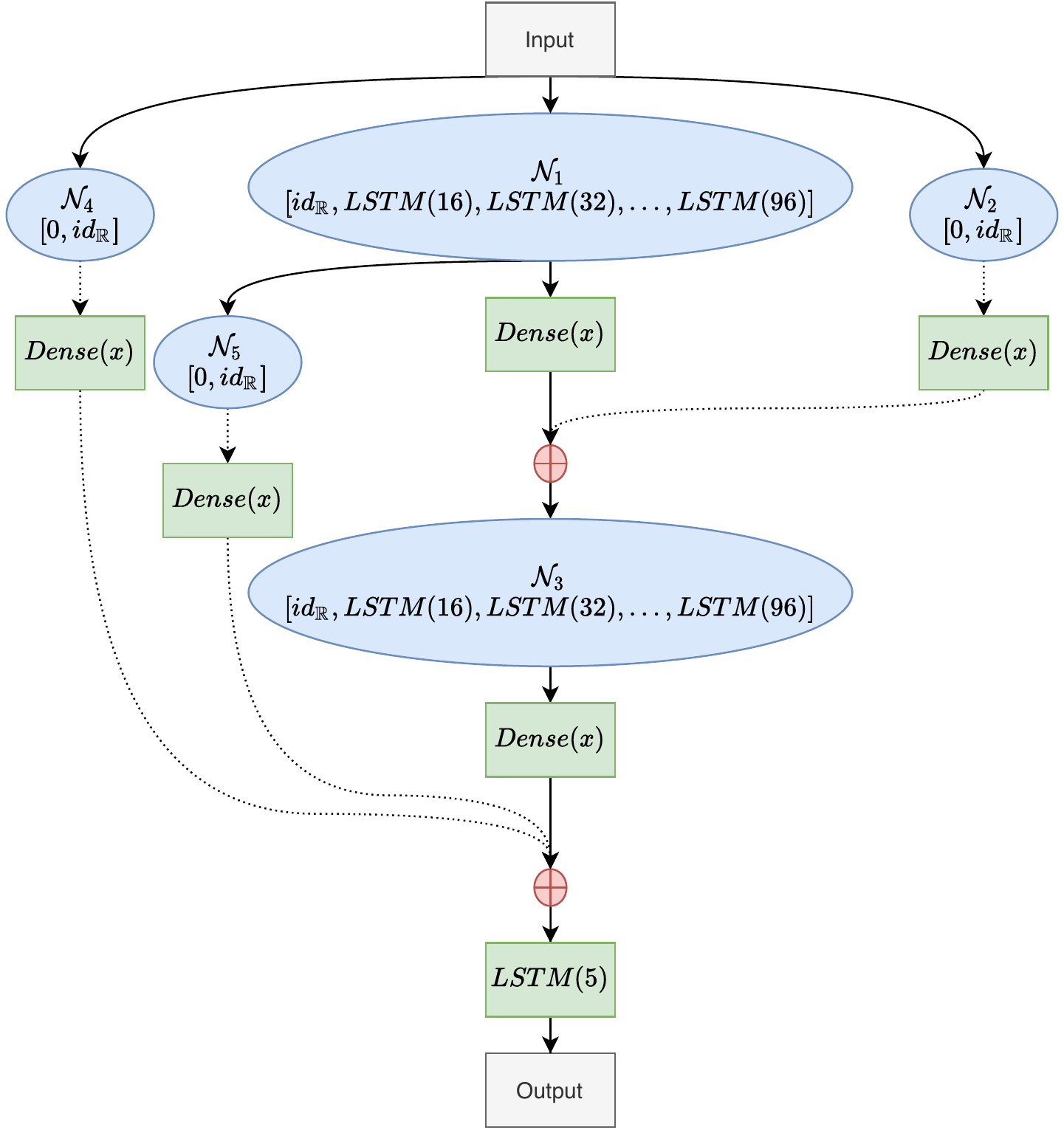}
    \caption{Example of a stacked LSTM search space for POD-LSTM with two variable LSTM nodes in blue, $\mathcal{N}_{1}$ and $\mathcal{N}_{3}$. The skip-connection variable nodes are $\mathcal{N}_{2}$, $\mathcal{N}_{4}$, and $\mathcal{N}_{5}$. Dotted lines represent possible skip connections. The last layer is a constant LSTM(5) node to match the output dimension of five.}
    \label{seach_space_schematic}
\end{figure}

In DeepHyper, the search space of the neural architecture is represented as a directed acyclic graph. The nodes representing inputs and outputs of the deep neural network are fixed and respectively denoted as $\mathcal{I}_i$ and $\mathcal{O}_j$. All other nodes $\mathcal{N}_k$ are called intermediate nodes, each with a list of operations (choices). Each intermediate node is a constant, a variable, or a skip connection node. A constant node's list contains only one operation, while a variable node's list contains multiple options. 
The formation of skip connections between the variable nodes is enabled by a skip-connection variable node. Given three nodes $\mathcal{N}_{k-1}, \mathcal{N}_{k}$, and $\mathcal{N}_{k+1}$, the skip connection node allows for the possible construction of a direct connection between $\mathcal{N}_{k-1}$ and $\mathcal{N}_{k+1}$. This skip connection node will have two operations, zero for no skip connection and identity for skip connection. In a skip connection, the tensor output from $\mathcal{N}_{k-1}$ is passed through a dense layer and a sum operator. Since the skip connections can be formed between variable nodes that have a different number of neurons, the dense layer is used to project the incoming tensor to the right shape when the skip connection is formed. The sum operator adds the two tensors from  $\mathcal{N}_{k-1}$ and $\mathcal{N}_k$ and passes the resulting tensor to $\mathcal{N}_{k+1}$. Without loss of generality, the same process can be followed for any number of nodes. For example, in the case of three nodes, two skip-connection nodes will be inserted before an incumbent node. For the stacked LSTM discovery, we define $m$ variable LSTM nodes, where each node is an LSTM layer with a list of numbers of neurons as possible operations. Figure \ref{seach_space_schematic} shows an example LSTM search space with two variable nodes. 

We note that the input and output nodes are determined by the shape of our training data and are immutable.  We also note  that the second dimension of a tensor that is transformed from input to output is kept constant for all experiments. This aligns with the temporal dimension of an LSTM and is not perturbed. 

\subsection{Algorithms for architecture discovery}

We use search methods in DeepHyper to choose from a set of possible integer values at each variable node. At the LSTM variable nodes, this choice decides the number of hidden layer neurons. At the skip connection variable nodes, this choice decides connections to previous layers of the architecture. For intelligently searching this space, we have implemented a recently introduced completely asynchronous evolutionary algorithm called aging evolution (AE) \cite{real2019regularized} within DeepHyper. In addition, DeepHyper supports two search methods for NAS: a parallel version of reinforcement learning (RL) \cite{balaprakash2019scalable, zoph2018learning} based on the proximal policy optimization \cite{schulman2017proximal} and a random search (RS). 

\subsubsection{Aging evolution} 

AE searches for new architectures by performing mutations without crossovers on existing architectures within a population. At the start of the search, a population of $p$ architectures is initialized randomly, and the fitness metric (for validation accuracy) is recorded. 
Following this initialization, samples of size $s$ are drawn randomly without replacement. A mutation is performed on the architecture with the highest accuracy within each sample (the parent) to obtain a new (child) architecture. A mutation corresponds to choosing a different operation for one variable node in the search space. This is achieved by first randomly sampling a variable node and then choosing (again at random) a value for that node excluding the current value. The validation accuracy of the child architecture is recorded. The child then is added to the population by replacing the oldest member of the population. For the purpose of mutation, an architecture is interpreted to be a sequence of integers, and certain probabilistic perturbations to this sequence are performed to obtain new architectures. Over multiple cycles, better architectures are obtained through repeated sampling and mutation. The sampling and mutation operations are inexpensive and can be performed quickly. When AE completes an evaluation, another architecture configuration for training is obtained by performing a mutation of the previously evaluated architectures (stored for the duration of the experiment in memory) and does not require any communication with other compute nodes.

\subsubsection{Distributed RL method}

RL is a framework where an agent interacts (or multiple agents interact) with an environment by performing actions and collecting rewards and observations from this same environment. In our case, actions correspond to operation choices for variable nodes in the NAS search space. The reward is the accuracy computed on the validation set. The RL method in DeepHyper uses proximal policy optimization \cite{schulman2017proximal} with a loss function of the form 
\begin{equation}
    J_t(\theta) =  \hat{\mathbb{E}}_t\lbrack \textrm{min}(r_t(\theta)\hat{A}_t, \textrm{clip}(r_t(\theta), 1-\epsilon, 1+\epsilon) \hat{A}_t \rbrack,
\end{equation}
where $r_t(\theta) = \frac{\pi_\theta(a_t|s_t)}{\pi_{\theta_{\rm old}}(a_t|s_t)}$ is the ratio of action probabilities under the new and old policies; the clip/median operator ensures that the ratio is in the interval $[1-\epsilon, 1+\epsilon]$; and $\epsilon \in (0,1)$ is a hyperparameter (typically set to 0.1 or 0.2).
The clipping operation helps stabilize gradient updates. The method adopts the multimaster-multiworker paradigm for parallelization. Each master runs a copy of a policy  and value neural network, termed an agent. It generates $b$ architectures and evaluates them in parallel using multiple worker nodes. 
The $b$ validation metrics are then collected by each agent to do a distributed computation of gradients.
The agents perform an all-reduce with the mean operator on gradients and use that to update the policy and value neural network. This procedure was chosen instead of asynchronous update because it has been empirically shown to  perform better (see, e.g., \cite{heess2017emergence}). 



\subsubsection{Random search method}

For  comparison of all our methods, we also describe results from a random search algorithm that explores the search space of architectures by randomly assigning operations at each node. This search is embarrassingly parallel: it does not need any internode communication.  The lack of an intelligent search, however, leads to architectures that do not improve over time or scale. Such results are  demonstrated in our experiments as well. 

\section{Experiments}

We used Theta, a 4,392-node, 11.69-petaflop Cray XC40–based supercomputer at the Argonne Leadership Computing Facility. Each node of Theta is a 64-core Intel Knights Landing processor with 16 GB of in-package memory, 192 GB of DDR4 memory, and a 128 GB SSD. The compute nodes are interconnected by an Aries fabric with a file system capacity of 10 PB. The software environment that we used consists of Python 3.6.6, TensorFlow 1.14 \cite{tensorflow2015-whitepaper} and DeepHyper 0.1.7. The NAS API within DeepHyper utilized Keras 2.3.1.

For the stacked LSTM search space, we used 5 LSTM variable nodes ($m=5$). This resulted in the creation of 9 skip connection variable nodes. The possible operations at each of the LSTM variable node were set to [Identity, LSTM(16), LSTM(32), LSTM(64), LSTM(80), LSTM(96)] representing different layer types: Identity layer, LSTM layer with 16, 32, 64, 80, and 96 units. The dense layers for projection did not have any activation function. After each add operation, the ReLU activation function was applied to the tensor. For this search space, the total number of architectures was 8,605,184.

As a default, we evaluated the three search methods on 128 nodes. For scaling experiments, we used different node counts: 33, 64, 256, and 512. We used 33 (instead of 32) nodes because in RL we set the number of agents to 11 and adapt the number of workers per agent based on the node count as prescribed in \cite{balaprakash2019scalable}. This implies that given any number of compute nodes, 11 are reserved solely for the agents whereas the rest function as workers that are equally distributed to each agent. With 33 nodes, each agent is assigned 2 workers, for a total of 33 compute nodes being utilized. When 64 nodes are used, each agent is allocated 4 workers, for a total of 55 used nodes and 9 unused nodes. Similarly, for 128 compute nodes, each agent is allocated 10 workers, for a total of 121 utilized and 7 unused nodes. For 256 and 512 compute nodes, each agent is provided 22 and 45 workers,  resulting in 3 and 6 unused nodes, respectively. The equal division of workers among agents is implemented for simplicity within DeepHyper. 
For each node count, each search method was run for 3 hours of wall time. 

Each evaluation in all three search methods involved training a generated network and returning the validation metric to the search method. The evaluation used only a single node (no multinode data-parallel training). The mean squared error was used for training the network, and the coefficient of determination ($R^2$) was used as a metric on the validation data set. The AE and RL methods were tasked with optimizing the $R^2$ metric by searching a large space of LSTM architectures. The RS method sampled configurations at random without any feedback. The training hyperparameters were kept the same in all of our experiments: batch size of 64,  learning rate of 0.001, and 20 epochs of training with the ADAM optimizer \cite{kingma2014adam}. We set the maximum depth of the stacked LSTM network (an upper bound held constant in all our searches) to 5.

To assess the efficiency of the search, we tracked the averaged reward (i.e., the validation accuracy of our architecture) with respect to wall-clock time.  To assess scaling, we recorded the averaged node utilization for each search; to assess the overall benefit, we selected the best  architecture found during the search and assessed it on the test data set. For this, we performed  posttraining, where the best-found architecture was retrained from scratch for a longer duration and tested on a completely unseen test data set. We note  that our metrics (given by the reward and node utilization) were computed by using a moving window average of window size 100.

\subsection{Comparison of different methods}

Here, we compare the three search methods and show that AE outperforms RL and RS by achieving a higher validation accuracy in a shorter wall-clock time. 

We ran AE, RL, and RS on 128 nodes on Theta. For the asynchronous algorithms of AE and RS, all nodes were considered workers because they are able to evaluate architectures independently of any master. For AE, we used a population size of 100 and a sample size of 10 to execute the mutation of architectures asynchronously. 

For RS, random configurations are sampled independently of any other nodes. In contrast, RL relies on a synchronous update where the compute nodes are divided into agents and workers. Each agent is given an equal number of workers to evaluate architectures and is provided rewards before the agent neural networks are updated by using the policy gradients. We fixed the number of agents for this and all subsequent experiments to 11. For a 128-compute node experiment with 11 agents, we had 10 workers per agent node, for a total of 110 workers and 11 agents; 7 nodes remained idle. 

\begin{figure}
    \centering
    \includegraphics[width=0.42\textwidth]{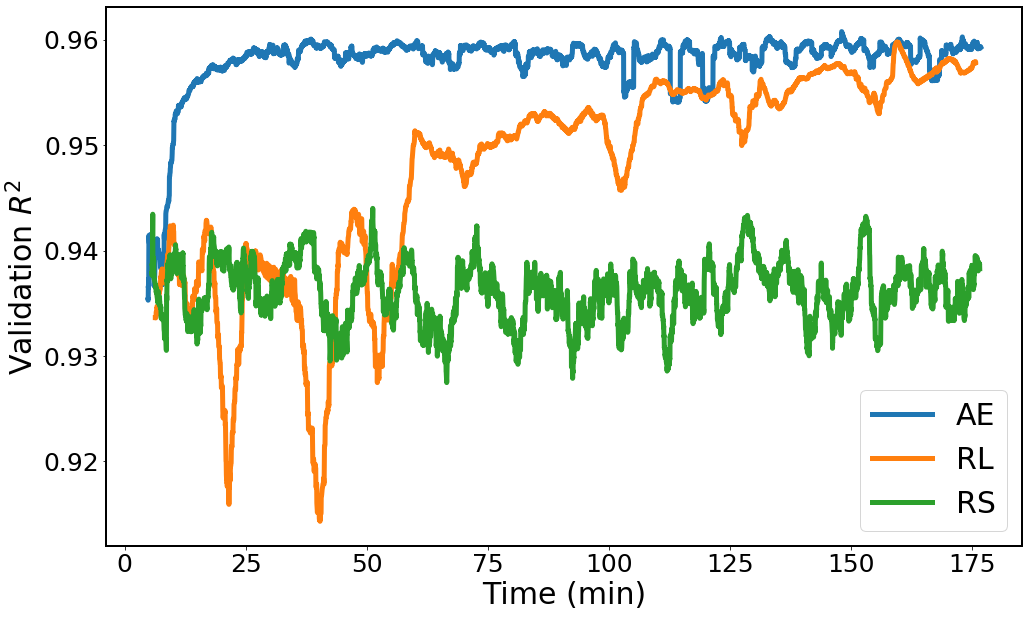}
    \caption{Comparison of search trajectories for AE, RL, and RS for 128 compute nodes on Theta. Each search was run for 3 hours of wall time. AE obtains optimal architectures in a much shorter duration. Both RL and AE are an improvement over random search methods.}
    \label{fig:128_rw}
\end{figure}

Figure \ref{fig:128_rw} shows the search trajectory of validation $R^2$ of our three search strategies with respect to wall-clock time. We observe that AE reaches a validation $R^2$ value of 0.96 within 50 minutes. On the other hand, RL exhibits strong exploration in the beginning of the search and reaches a $R^2$ value comparable to that of AE at 160 minutes. The RS without any feedback mechanism finds architectures with $R^2$ values between 0.93 and 0.94. The results of RS show the importance of having a feedback-based search such as AE and RL.

The superior performance of AE can be attributed to its effective aging mechanism and the resulting regularization, as discussed in \cite{real2019regularized}. In AE, the individuals in the population die faster; an architecture can stay alive for a long time only through inheritance from parent to child for a number of generations. When that number is passed, the architecture undergoes retraining; and if the retraining accuracy is not high, the architecture is removed from the population. An architecture can remain in the population for a long time only when its retraining accuracy is high for multiple generations. Consequently, the aging mechanism helps navigate the training noise in the search process and provides a regularization mechanism. RL lacks such a regularization mechanism, and the slower convergence can be attributed to the synchronous gradient update mechanism at the inter- and intra-agent levels. 



\subsection{Posttraining and science results}
\label{science-results}

To ensure efficient NAS, one commonly solves a smaller problem during the search itself before utilizing the best architecture discovered for the larger training task. This approach helps us explore the large space of neural architectures more efficiently. It also helps us deal with larger data sets if needed. In this study, we utilized fewer epochs of training (20 epochs) during the search and retrained from scratch using a greater number of epochs during posttraining (100 epochs). Here, we show that retraining the best-stacked LSTM architecture obtained from AE results in significant improvement. \revised{This phase is called posttraining to differentiate it from the commonly used augmentation phase of many NAS algorithms where the best-architecture is augmented with additional layers and retrained on the full problem}.

\begin{figure}
    \centering
    \includegraphics[width=0.2\textwidth]{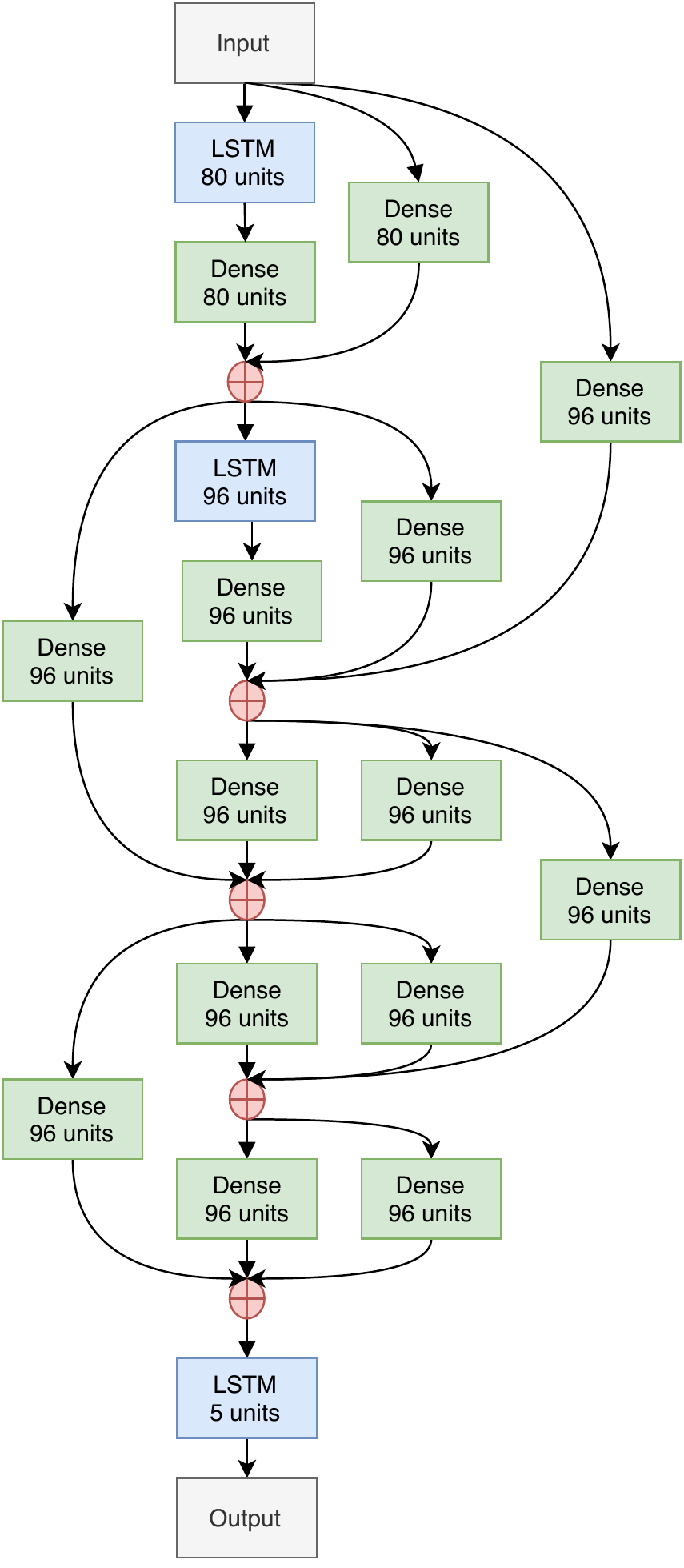}
    \caption{Best-found LSTM architecture for the NOAA SST data set using the aging evolution search strategy on 128 compute nodes of Theta for 3 hours of wall time.}
    \label{fig:networks}
\end{figure}

Figure \ref{fig:networks}  shows the best architecture found by AE with 128 nodes. One can observe the unusual nature of our network as evidenced by multiple skip connections. We utilized the best architecture found by AE (in terms of validation $R^2$) for posttraining and scientific assessments.

\revised{For posttraining, we used the same hyperparameters as specified in the NAS with the exception of a longer training duration of 100 epochs (instead of 20 for the search). A final validation $R^2$ value of 0.985 was observed, and this trained architecture was used for our science assessments.} Our architecture search as well as our posttraining utilized a sequence-to-sequence learning task where the historical temperatures (in a sequence) were used to predict a forecast sequence of the same length (i.e., measurements of 8 weeks of sea-surface temperature data were utilized to predict 8 weeks of the same in the future). This may also be seen in the output space of the best-found architectures where the second dimension of the output tensor is the same as the one used for the input.

\begin{figure*}
    \centering
    \mbox{
    \subfigure[NOAA SST training data forecast]{\includegraphics[width=0.48\textwidth]{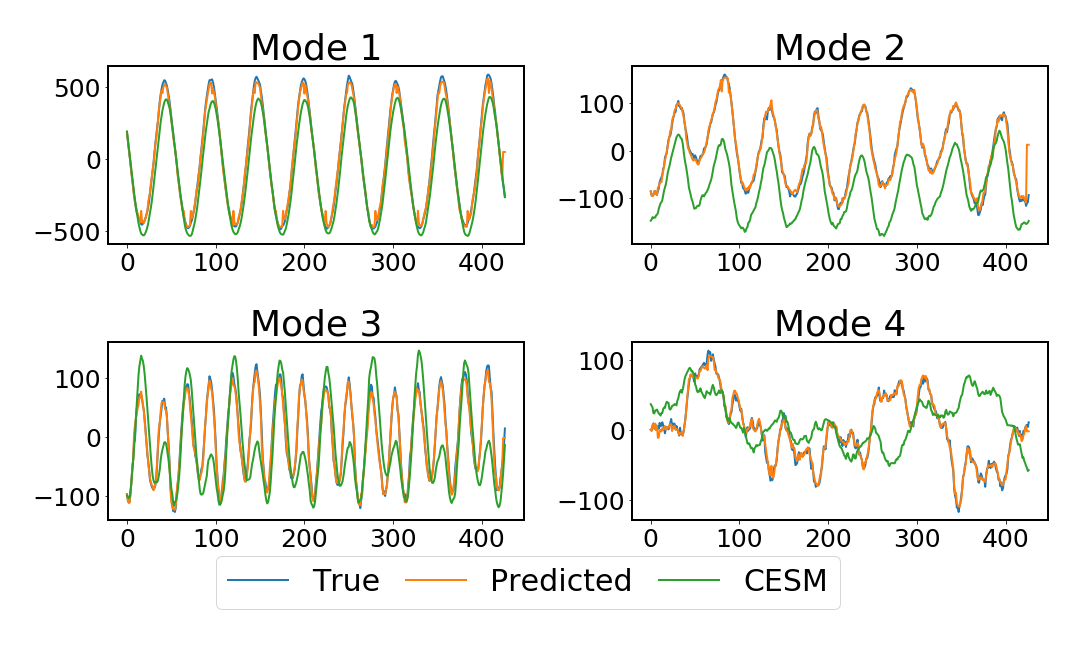}}
    \subfigure[NOAA SST testing data forecast]{\includegraphics[width=0.48\textwidth]{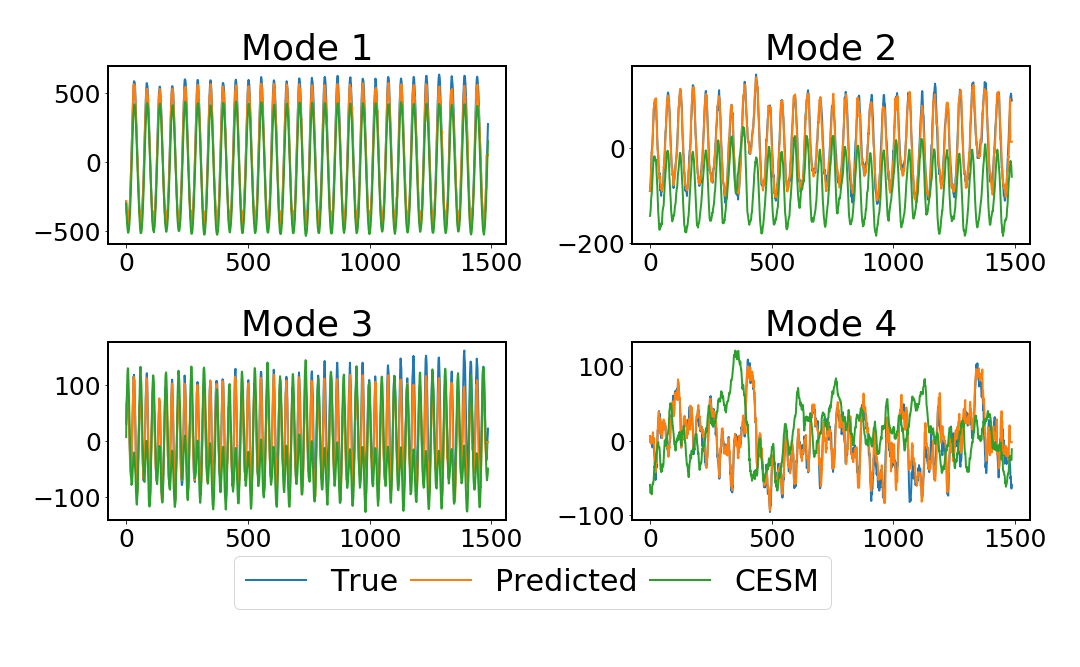}}
    } \\
    
    \caption{Posttraining results with progress to convergence for the optimal architecture showing improved performance for longer training durations (top row). Training and testing forecasts for the NOAA SST data set (bottom row) display the a posteriori performance after surrogate training. Comparisons with CESM show that higher frequencies may be predicted more accurately for short-term forecasts using the proposed method.}
    \label{fig:post_training}
\end{figure*}

Figure \ref{fig:post_training} shows the forecasts for the POD coefficients on  both the training and the testing data sets. While the training data (representing temperatures between 1981 and 1989)  predicts very well, the test data (from 1990 to 2018) shows a gradual increase in errors, with  later data points  far away from the training regime. In this figure, the modes refer to the linear coefficients that premultiply the POD basis vectors prior to reconstruction in the physical space. They may be interpreted to be the contribution of the different basis vectors (and correspondingly the different spatial frequency contributions) in the evolving flow. One can also observe that the amount of stochasticity increases significantly as the modal number increases. This increase may be explained by the fact that while seasonal dynamics are cyclical and lead to repeating patterns at the global scale (mode 1, 2, and 3), small-scale fluctuations may be exceedingly stochastic (mode 4 and beyond). We note that there are no external inputs to our data set and that the networks are utilized for forecasting under the assumption that the past information is always 100\% accurate. Simply speaking, the outputs of the LSTM forecast are not reused as inputs for future forecasts. The past is always known a priori. 

\revised{We also performed comparisons with data extracted (within the appropriate time range) from the Community Earth System Model (CESM) \cite{kay2015community}. The CESM forecasts represent a state-of-the-art process-based climate modeling system based on coupled numerical PDE evaluations for atmospheric, oceanic, land carbon cycle, and sea-ice component models. These are in addition to diagnostic biogeochemistry calculations for the oceanic ecosystem and the atmospheric carbon dioxide cycle. The CESM forecast data is made publicly available for climatologists because of the large compute costs incurred in its simulation\footnote{http://www.cesm.ucar.edu/projects/community-projects/LENS/data-sets.html}. Figure \ref{fig:post_training} also shows the coefficients of the CESM data projected onto the NOAA POD modes. We note that the POD coefficients of the CESM forecasts tend to pick up trends in the large-scale features (i.e., modes 1 and 2) appropriately but show distinct misalignment with increasing modes. Another important fact is that the CESM model forecasts were performed on a finer grid (320 $\times$ 384 degrees of freedom for oceanic regions alone) and some errors may be due to cubic interpolation onto the remote sensing grid. We also note that the CESM forecasts are based on specifying one initial condition and running climate simulations for multiple coupled geophysical processes for centuries. In contrast, the POD-LSTM methodology detailed in this paper relies on a short (8-week) forecast given \emph{true} observations from the past. Therefore, we stress that the proposed emulation strategy and the long-term climate forecasts using PDEs are designed for different use cases. Although POD-LSTM provides more accurate estimates, CESM may provide better results for longer forecast horizons.}

\revised{To perform a fairer assessment of our proposed framework, we also tested our predictions against forecasts from the Global Hybrid Coordinate Ocean Model (HYCOM)---the current state-of-the-art SST forecast technique.\footnote{https://www.ncdc.noaa.gov/data-access/model-data/model-datasets/navoceano-hycom-glb} In contrast to CESM, HYCOM provides short-term 4-day forecasts at 3-hour time steps, updated daily, and relies on a 1/12 degree grid (i.e., 144 times finer than the NOAA grid). We used aggregate HYCOM data (currently available only between April 5, 2015, and June 24, 2018) for a comparison with our proposed model emulator. The extracted HYCOM data was interpolated onto the NOAA grid-coordinates before an assessment of root mean square errors (RMSEs). In particular, we focus on RMSEs in the Eastern Pacific region (between -10 to +10 degrees latitude and 200 to 250 degrees longitude) because of the importance of predicting abrupt SST rises in this region as well its distance from landmasses that may contribute to interpolation errors.  In addition, we provide a weekly breakdown of metrics within this time period and also show comparisons with CESM in Table \ref{RMSE_Table}. The trained POD-LSTM is seen to be competitive with both process-based models. However, we highlight the fact that the slightly larger biases in the CESM and HYCOM data may be an artifact of interpolation from grids of different spatiotemporal resolution. In either case, the POD-LSTM represents a viable opportunity for accelerated forecasting without the traditional constraints of either system of PDEs. }

\begin{table*}[]
\centering
\caption{RMSE breakdown (in Celsius) for different forecast techniques compared against the NAS-POD-LSTM forecasts between April 5, 2015, and June 24, 2018, in the Eastern Pacific region (between -10 to +10 degrees latitude and 200 to 250 degrees longitude). The proposed emulator  matches the accuracy of the process-based models for this particular metric and assessment.}
\begin{tabular}{|c|c|c|c|c|c|c|c|c|}
\hline
\multicolumn{1}{|c|}{} & \multicolumn{8}{c|}{RMSE ($^\circ$Celsius) }\\
\hline
 & Week 1 & Week 2 & Week 3 & Week 4 & Week 5 & Week 6 & Week 7 & Week 8 \\ \hline
Predicted & 0.62 & 0.63 & 0.64 & 0.66 & 0.63 & 0.66 & 0.69 & 0.65  \\ \hline
CESM      & 1.88 & 1.87 & 1.83 & 1.85 & 1.86 & 1.87 & 1.86 & 1.83  \\ \hline
HYCOM     & 0.99 & 0.99 & 1.03 & 1.04 & 1.02 & 1.05 & 1.03 & 1.05 \\ \hline
\end{tabular}
\label{RMSE_Table}
\end{table*}

\begin{figure*}
    \centering
    \subfigure[NOAA (Truth)]{\includegraphics[width=0.24\textwidth]{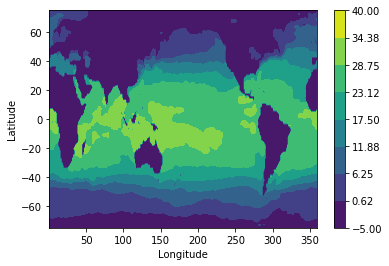}}
    \subfigure[HYCOM]{\includegraphics[width=0.24\textwidth]{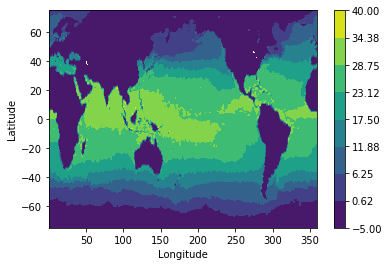}}
    \subfigure[CESM]{\includegraphics[width=0.24\textwidth]{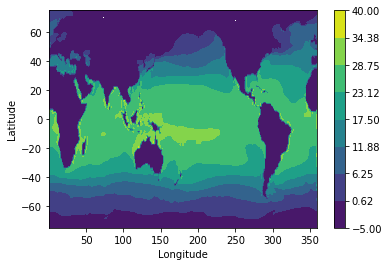}}
    \subfigure[POD-LSTM]{\includegraphics[width=0.24\textwidth]{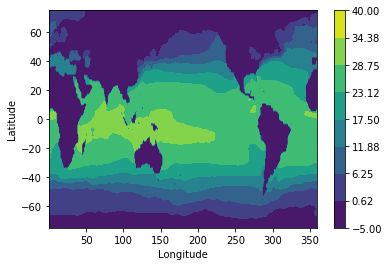}}
    \caption{Sample averaged temperature forecasts in degrees Celsius for the week starting  June 14, 2015 (within the testing regime for our machine learning framework). Note that CESM forecast horizons span several decades whereas HYCOM provides 4-day short-term forecasts.}
    \label{fig:NOAA_contour}
\end{figure*}

\begin{figure*}
    \centering
    \mbox{
    \subfigure[-5 $^{\circ}$ latitude, 210 $^{\circ}$ longitude]{\includegraphics[width=0.32\textwidth]{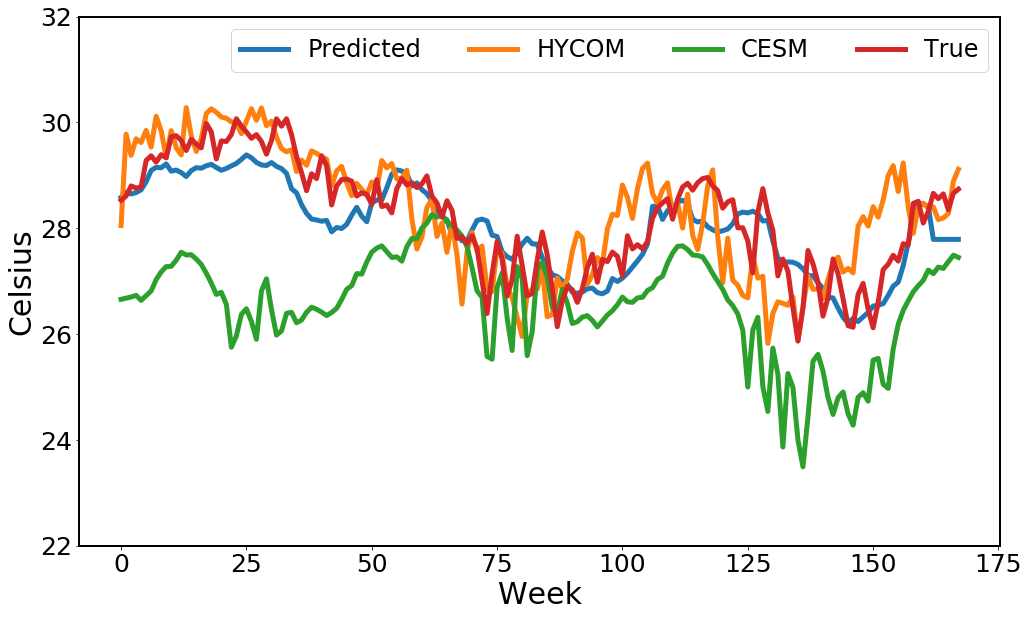}}
    \subfigure[+5 $^{\circ}$ latitude, 250 $^{\circ}$ longitude]{\includegraphics[width=0.32\textwidth]{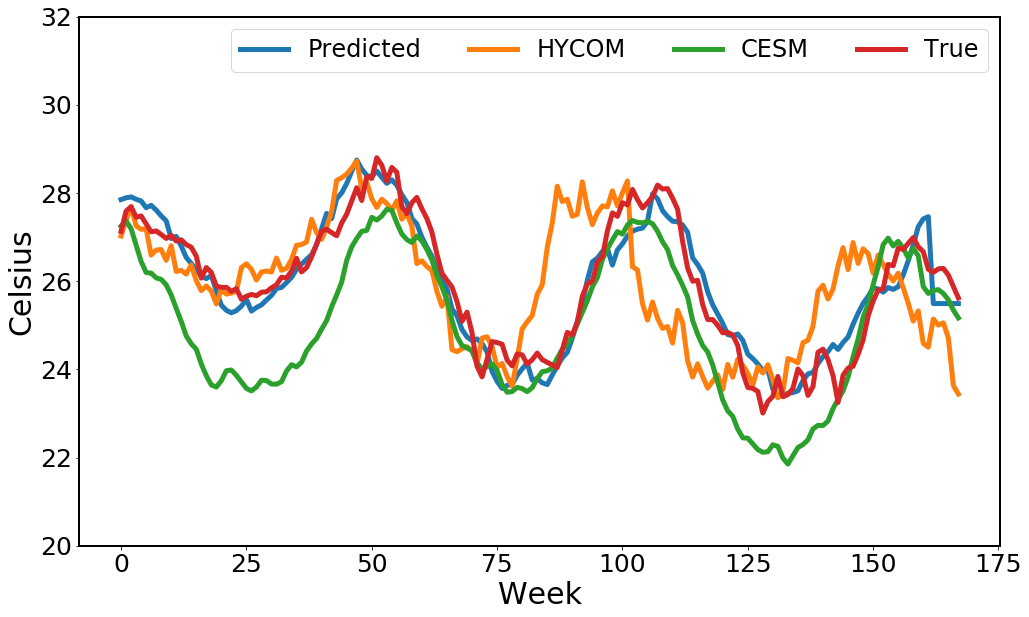}}
    \subfigure[+10 $^{\circ}$ latitude, 230 $^{\circ}$ longitude]{\includegraphics[width=0.32\textwidth]{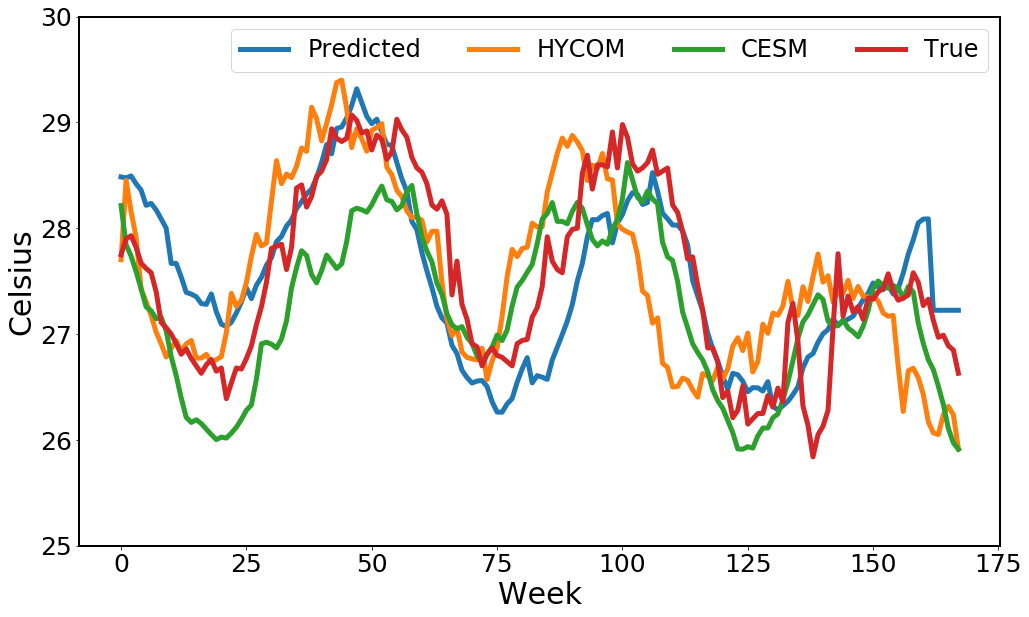}}
    }
    \caption{Temporal probes for the temperature at three discrete locations within the Eastern Pacific zone. HYCOM and POD-LSTM are shown to perform equally well while CESM makes slight errors in forecast due to its long-term forecast formulation. The data are plotted for weeks between  April 5, 2015, and June 17,  2018.}
    \label{fig:NOAA_probe}
\end{figure*}

\revised{An example forecast within the testing range (for the week starting June 14, 2015) is shown in Figure \ref{fig:NOAA_contour}, where the larger structures in the temperature field are captured effectively by the emulator. We note here that the POD-LSTM framework is fundamentally limited by the fact that the spectral content of the predictions can at best match the spectral support of the number of POD modes retained. Thus, we may interpret this to be training and forecasting on a filtered version of the true data set, where the truncation of the POD components leads to limited recovery of high-frequency information. We also observe that CESM and HYCOM predictions show qualitative agreement with the true data at larger scales. We show point estimates for forecasts from the different process-based frameworks at different locations in the Eastern Pacific ocean in Figure \ref{fig:NOAA_probe}, which show good agreement between HYCOM and POD-LSTM. CESM, as expected, is slightly inaccurate for short-term time scales because of its formulation. Both HYCOM and POD-LSTM capture seasonal trends appropriately within our testing period.}



\subsection{Comparisons with baseline machine learning models}

Here, we compare our automatically generated POD-LSTM network with manually generated POD-LSTM networks and with other classical forecasting methods and show the efficacy of our approach.

The classical forecasting assessments use linear, XGBoost, and random forest models within the non-autoregressive framework with no exogenous inputs; in other words, models are fit between an input space corresponding to a historical sequence of temperature to forecast the next sequence. These methods are all deployed within the \texttt{fireTS}\footnote{https://github.com/jxx123/fireTS} package;   details may be found in \cite{xie2018benchmark}. Briefly, if our target is given by $\mathbf{a}(t+1), \mathbf{a}(t+2), \hdots,  \mathbf{a}(t+K)$, we fit a data-driven regressor using information from $\mathbf{a}(t-1), \mathbf{a}(t-2), \hdots, \mathbf{a}(t-K)$, where $K$ is now interpreted to be the autoregression order. \revised{A prediction is made by using the underlying regressor while assuming that the past information is always coming from the true measurements. Note that all our baseline regression frameworks were implemented through the scikit-learn package \cite{sklearn} with default configurations}. For our manually designed LSTMs, we developed simple stacked architectures and scanned across the number of hidden layer neurons for each cell. Our LSTMs utilized both one and five hidden layers and demonstrated the challenge of manual model selection. These LSTMs also utilized 100 epochs for training. 

Table \ref{Table_Methods} shows the results. The $R^2$ metrics show that the best architecture found by DeepHyper (denoted NAS-POD-LSTM) outperforms the manually designed LSTMs and classical time-series prediction methods, with the highest $R^2$ value of 0.876. In particular, the benefit of using LSTMs over their counterparts is easily observed where LSTMs are seen to be more accurate ($R^2>0.73$) than the linear model ($R^2 \approx 0.1$), XGBoost ($R^2 \approx -0.1$) and the random forest regressor ($R^2 \approx 0.0$). 

\begin{table*}[]
\centering
\caption{Coefficients of determination ($R^2$) of different data-driven forecasting methods on the NOAA SST data set. Here, training data and validation data are obtained from 1981 to 1989, whereas testing data is obtained from 1990 to 2018. Note that the LSTM metrics are expressed in one-layered/five-layered configuration.}
\begin{tabular}{|c|c|c|c|c|c|c|c|c|}
\hline
\textbf{Model}     & \textbf{NAS-POD-LSTM} & \textbf{Linear} & \textbf{XGBoost} & \textbf{Random Forest} & \textbf{LSTM-40} & \textbf{LSTM-80} & \textbf{LSTM-120} & \textbf{LSTM-200} \\ \hline
\textbf{1981-1989} & \textbf{0.985}                 & 0.801           & 0.966            & 0.823                  & 0.916/0.944      & 0.931/0.948      & 0.922/0.956       & 0.902/0.963       \\ \hline
\textbf{1990-2018} & \textbf{0.876}                 & 0.172           & -0.056           & 0.002                  & 0.742/0.687      & 0.734/0.687      & 0.746/0.711       & 0.739/0.724       \\ \hline
\end{tabular}
\label{Table_Methods}
\end{table*}

In terms of times to solution, all data-driven models (i.e., models based around the forecast of POD coefficients) provided forecasts for the given time period (1981--2018) almost instantaneously. For instance, the NAS-LSTM model requires approximately 10 seconds to make forecasts in POD coefficient space from which entire fields can be reconstructed instantaneously by using the linear reconstruction operation. \revised{This can be contrasted with CESM (for the forecast period of 1920--2100), which required 17 million core-hours on Yellowstone, NCAR's high-performance computing resource, for each of the 30 members of the large ensemble. While compute costs in such high detail are not available for HYCOM, this short-term ocean prediction system runs daily at the Navy DoD Supercomputing Resource Center, with daily data typically accessible within 48 hours of the initial run time\footnote{https://www.hycom.org/dataserver}. Benchmarking for the 1/25 degree HYCOM forecasts (twice finer than the reference data used here) indicates the requirement of 800 core-hours per day of forecast on a Cray XC40 system.\footnote{https://www.hycom.org/attachments/066\_talk\_COAPS\_17a.pdf}}


\subsection{Scaling}
Here, we study the scaling behavior of the three search methods. We show that AE scales better than RL and outperforms RS with respect to accuracy. In addition to 128 nodes, we utilized 33, 64, 256, and 512 nodes for the scaling experiments. We analyzed the scaling with respect to node utilization, number of architectures evaluated, and number of high-performing architectures.


\noindent {\bf Node utilization:} Table \ref{Scaling_table} shows the average node utilization of the search methods for different node counts. The metric used in this table is computed by a ratio for the observed effective node utilization of an experiment against the ideal node utilization (for AE, RS, and RL). The observed effective node utilization over 3 hours of wall time is computed by using an area-under-the-curve (AUC) calculation. Subsequently, this area is divided by the ideal node utilization AUC. Therefore, a metric closer to 1 implies optimal utilization of all compute nodes. We note that the trapezoidal integration rule is utilized for the AUC calculation. We observe that the node utilization of AE and that of RS are similar and are above 0.9 for up to 256 nodes; for 512 nodes, the utilization drops below 0.87. \revised{The node utilization for RL is poor, hovering around 0.5 for the different compute node experiments. This can be attributed to two factors: the gradient averaging across agents is synchronous, and each agent needs to wait until all its workers finish their evaluations before computing the agent-specific gradient. In the beginning, each agent will generate a wide range of architectures, and each can have a different training time. The worker nodes for a given agent cannot proceed to the next batch of evaluations, and they become idle because  one or more worker nodes require more time to finish the training. This situation has previously been observed in \cite{balaprakash2019scalable} within DeepHyper. Note that AE and RS do not have such utilization bottlenecks.}

\revised{\noindent {\bf Number of architectures evaluated:} We observe that AE consistently is able to evaluate more architecture evaluations than RS and RL can for a fixed wall time (see Table \ref{Scaling_table}). The advantage over RL may be explained by the fact that RL requires a synchronization before obtaining rewards for a batch of architectures (also evidenced by lower node utilization). The improvement over RS may be attributed to the AE algorithm prioritizing architectures with fewer trainable parameters---a fact that has been demonstrated for asynchronous algorithms previously \cite{balaprakash2019scalable}. When 33 compute nodes are used, the AE strategy  completes 2,093 evaluations compared with only 1,066 by RL and 1,780 by RS. For 64 compute nodes, the total number of evaluations for AE increases significantly, with 4,201 evaluations performed successfully, whereas RL and RS lead to 2,100 and 3,630 evaluations in the same duration, respectively. As mentioned previously, for 128 nodes, AE performs 8,068 evaluations whereas RL and RS evaluate 4,740 and 7,267 architectures, respectively. For 256 nodes, AE, RL, and RS evaluate 18,39, 9,680, and 15,221 architectures, respectively. Similar trends are seen for 512 nodes, with 33,748, 16,335, and 26,559 architectures evaluated by AE, RL, and RS, respectively. We note that the AE strategy evaluates roughly double the number of architectures for all compute node experiments in comparison with RL.} 

\begin{figure*}
    \centering
    \subfigure[AE-discovered architectures: Temporal breakdown]{\label{subfig:thresh_sf_a}\includegraphics[width=0.42\textwidth]{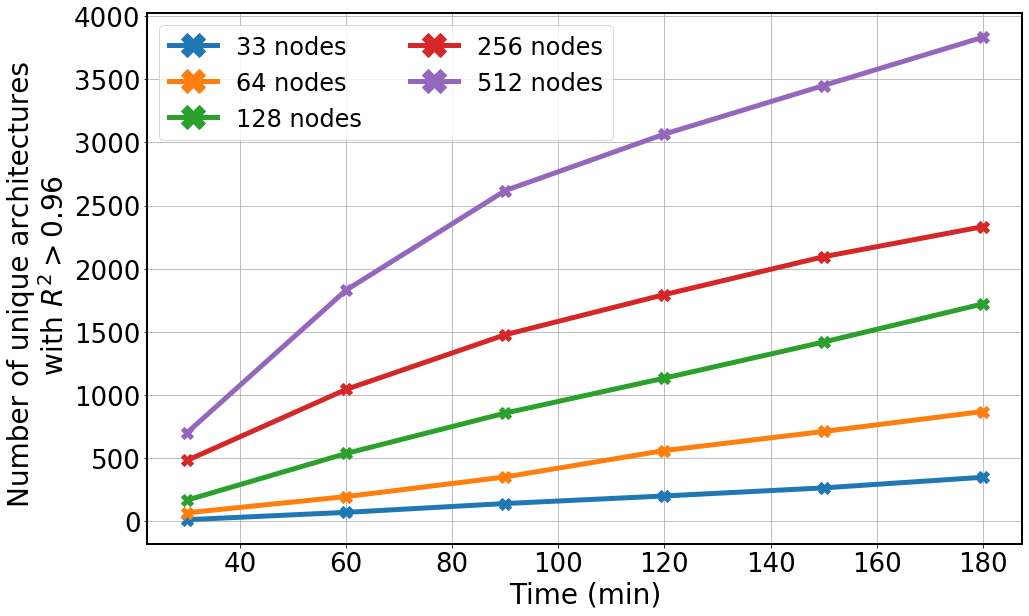}}
    \subfigure[High-performing architectures ]{\label{subfig:thresh_sf_b}\includegraphics[width=0.42\textwidth]{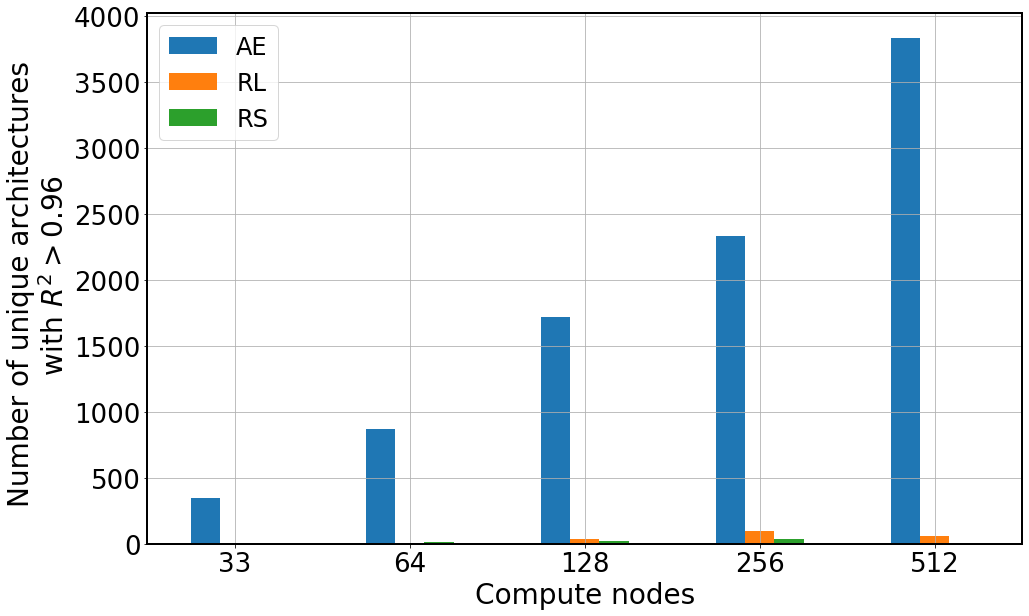}}
    \caption{Percentage of architectures with $R^2$ greater than $0.96$ for different compute nodes. The plot on the left shows that the AE search strategy is more effective at obtaining better architectures.}
    \label{fig:threshold_comp}
\end{figure*}

\revised{\noindent {\bf High-performing architectures discovered:}
Since AE, RL, and RS are randomized search methods, ideally one would need multiple runs with different random seeds to exactly assess the impact of scalability. However, running multiple repetitions is computationally infeasible: for instance, 10 repetitions requires 450 hours (=$10$ runs $\times$ $3$ methods $\times$ $5$ node counts $\times$ $3$ hours) if we run each job sequentially. Therefore, we analyzed the high-performing architecture metric,  defined as the number of unique discovered architectures that have $R^2$ greater than $0.96$. Figure \ref{subfig:thresh_sf_a} shows this metric as a function of time for AE. We can observe that the number of unique architectures obtained by AE grows considerably with greater numbers of compute nodes. In particular, the number of such architectures obtained by AE at 180 minutes using 33 nodes is achieved by AE with 64 nodes in 90 minutes, and the cumulative number of unique architectures at the end of the search is much higher as well. We can observe a similar trend as we double the node counts: in 90 minutes, 128 nodes obtains a number of unique architectures similar to that of  the entire 64-node run. The 256-node search obtains a number of unique architectures in 120 minutes similar to that of the entire 128-node run, and the 512-node search finds more unique architectures in 90 minutes than in the entire 256-node search. These results suggest that the AE algorithm is well suited to discovering unique high-performing neural architectures from a large search space on up to at least 512 KNL nodes.}

\revised{
For a thorough comparison of AE with the other algorithms, we also examined the number of unique architectures obtained at the end of all 180-minute searches. The results are shown in Figure \ref{subfig:thresh_sf_a}. We can see that the AE method outperforms the RL and RS strategies comprehensively. Noticeably, the number of unique architectures discovered by RL is seen to saturate after 256 nodes, indicating possible issues with scaling these types of synchronous algorithms on KNL nodes. Given these results, we recommend the use of AE for the most efficient exploration of a given search space on KNL architectures.}


\begin{table}
\centering
\caption{Tabulation of node utilization and total number of evaluations for different search strategies on varying numbers of compute nodes of Theta.}
\begin{tabular}{|c|c|c|c||c|c|c|}
\hline
                      & \multicolumn{3}{c||}{\textbf{Node utilization}} & \multicolumn{3}{c|}{\textbf{Number of evaluations}} \\ \hline
\textbf{No. of nodes} & \textbf{AE}    & \textbf{RL}   & \textbf{RS}   & \textbf{AE}  & \textbf{RL}  & \textbf{RS} \\ \hline
\textbf{33}           & 0.905          & 0.592         & \textbf{0.913}         & \textbf{2,093}         & 1066         & 1780   \\ \hline
\textbf{64}           & 0.920          & 0.482         & \textbf{0.927}         & \textbf{4,201}         & 2100         & 3630  \\ \hline
\textbf{128}          & 0.918          & 0.527         & \textbf{0.921}         & \textbf{8,068}         & 4740         & 7267  \\ \hline
\textbf{256}          & 0.911          & 0.509         & \textbf{0.936}         & \textbf{18,039}        & 9680         & 15221 \\ \hline
\textbf{512}          & \textbf{0.962}          & 0.541            & 0.869         & \textbf{33,748}        & 16335           & 26559 \\ \hline
\end{tabular}
\label{Scaling_table}
\end{table}

\subsection{Variability analysis}

\begin{figure*}
    \centering
    \mbox{
    \subfigure[Averaged reward - AE]{\includegraphics[width=0.35\textwidth]{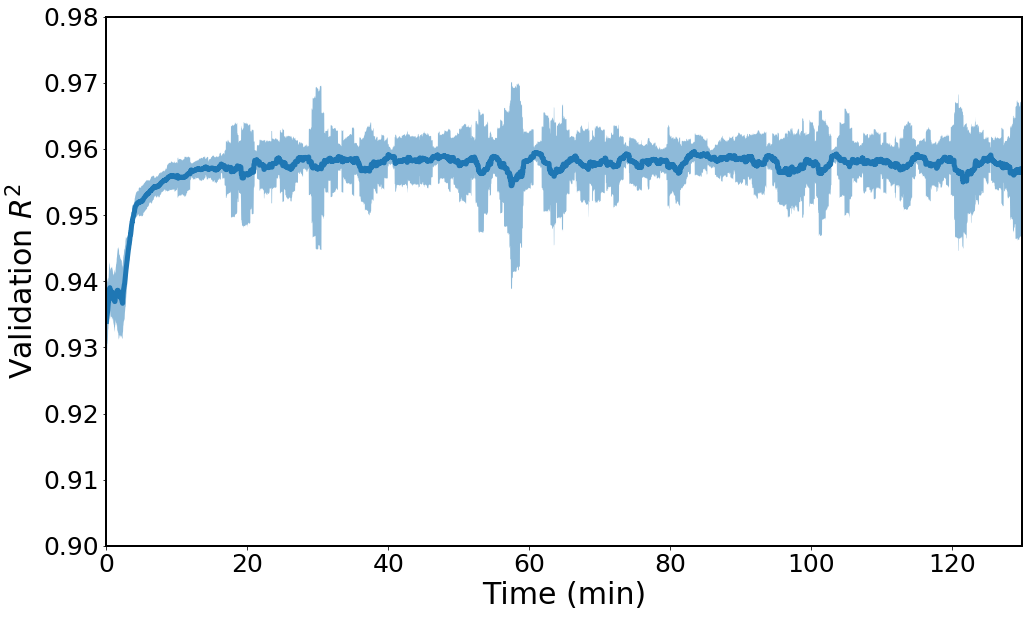}}
    \subfigure[Node utilization - AE]{\includegraphics[width=0.35\textwidth]{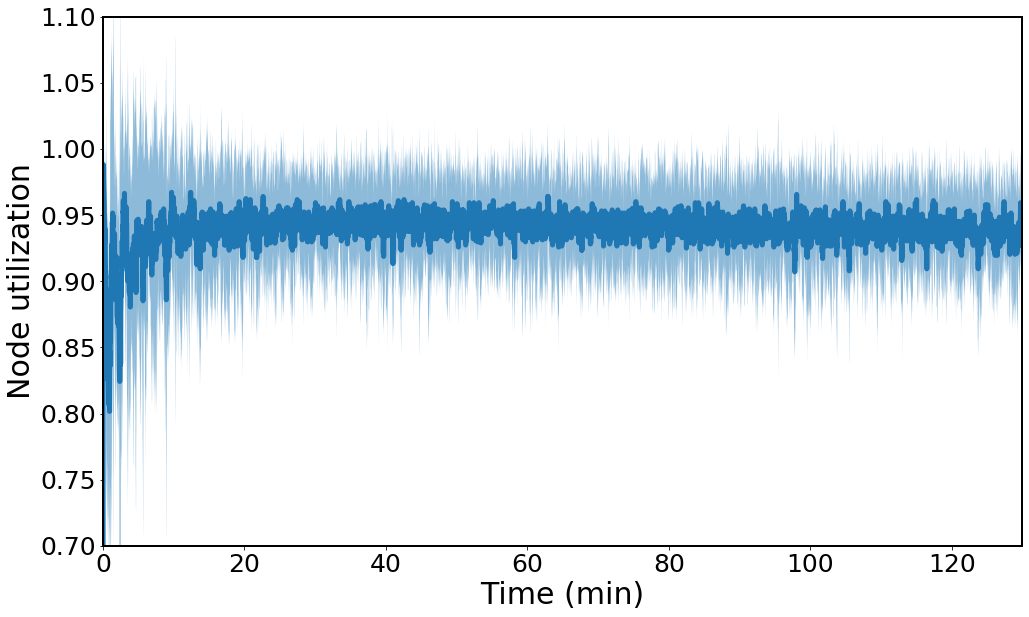}}
    } \\
    \mbox{
    \subfigure[Averaged reward - RL]{\includegraphics[width=0.35\textwidth]{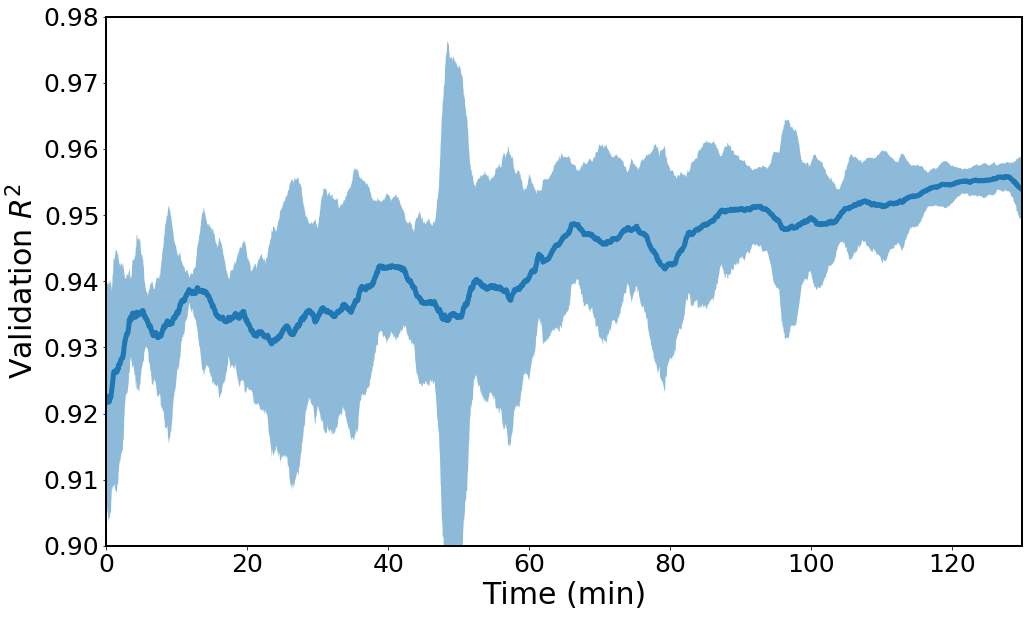}}
    \subfigure[Node utilization - RL]{\includegraphics[width=0.35\textwidth]{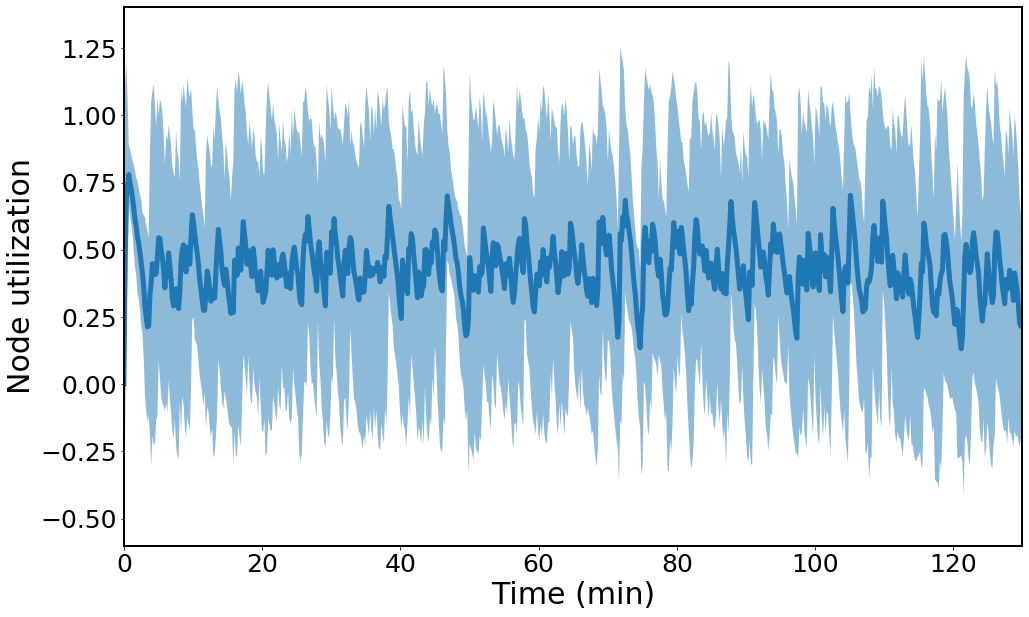}}
    }
    \caption{Mean and two standard-deviation-based confidence intervals for 10 NAS evaluations for AE (top) and RL (bottom) on 128 nodes. The low variability of AE indicates that the optimal performance of this search algorithm was not fortuitous. The oscillatory behavior in node utilization for RL can be observed across different experiments.}
    \label{fig:evo_ppo_bounds}
\end{figure*}

Our previous set of experiments shows that AE strikes the right balance between optimal node utilization and reward-driven search by avoiding the synchronization requirements of RL. To ensure that the behavior of AE is repeatable, we made 10 runs of AE, each with different random seeds, on 128 nodes. The results are shown in Figure \ref{fig:evo_ppo_bounds}. The averaged reward and node utilization curves are shown by their mean value and by shading of two standard deviations obtained from 10 runs. We also show results from performing 10 experiments using RL, where the oscillatory node utilization is replicated for different random seeds. We also replicate the slower growth of the reward as observed previously in Fig. \ref{fig:128_rw}. Overall, the values reflect the performance observed in the previous run that used AE on 128 nodes.

\section{Related work}

POD-based compression coupled with LSTM-based forecasting has been utilized for multiple physical phenomena that suffer from numerical instability and imprecise knowledge of physics. For instance, the POD-LSTM method has been used for building implicit closure models for high-dimensional systems \cite{maulik2020time,rahman2019nonintrusive}, for turbulent flow control \cite{mohan2018deep}, and for exogenous corrections to dynamical systems \cite{ahmed2020long}. Our work is the first deployment of the POD-LSTM method for real-world geophysical data forecasting. There have been recent results from the use of convolutional LSTMs (i.e., using a nonlinear generalization of POD) for forecasting on computational fluid dynamics data \cite{xingjian2015convolutional,mohan2019compressed}. However, we have performed this study using POD because of the spectrally relevant information associated with each coefficient of the forecast. More important, previous POD-LSTMs were manually designed, whereas we have demonstrated NAS for automated POD-LSTM development, a significant improvement in the state of the art. 

Several studies of automated recurrent network architecture search have been conducted for a diverse set of applications. One of the earliest examples that is similar to our approach was GNARL  \cite{angeline1994evolutionary}, which utilized evolutionary programming to search for the connections and weights of a neural network with applications to natural language processing. The approach was deployed serially, however, and relied on a synchronous update strategy where 50\% of the top-performing architectures are retained while the rest are discarded at each cycle. More recently, there has been research into the use of differentiable architecture search \cite{DBLP:journals/corr/abs-1806-09055}, where a continuous relaxation of the discrete search space was used to instantiate a hyper-neural network including all possible models. Based on that, a bilevel stochastic-gradient-based optimization problem was formulated. Nevertheless, this method does not show stable behavior when using skip connections because of unbalanced optimization between different possible operations at variable nodes. Evolutionary NAS has also been utilized to discover the internal configurations of an LSTM cell itself; for instance, in \cite{rawal2018nodes} a tree-based encoding of the LSTM cell components was explored for natural language and music modeling tasks. Other studies have looked into optimizing parameters that control the number of loops between layers of a network during the stochastic-gradient-based learning process \cite{savarese2019learning} for image classification applications. This approach leads to recurrent neural architecture discovery during the learning process itself. In contrast, in our study, architectures are generated and evaluated separately at scale. 

Evolutionary-algorithm-based architecture searches have also been deployed at scale, for example in \cite{ororbia2019investigating}, where hybridizations of LSTM cells as well as simpler nodes were studied by using neuroevolution \cite{stanley2002evolving}. Notably, this work was deployed at scale and also dealt with forecasting of real-world data obtained from the aviation and power sectors. Our approach is also able to integrate hybridizations of fully connected, skip, and identity layers in its search space at scale. A recent investigation in \cite{huang2019wenet} also poses the recurrent NAS problem as the determination of weighting parameters for different networks that act as a mixture of experts applied to language modeling tasks. Our work differs from a majority of these investigations  by combining the use of scale for architecture discovery for forecasting on a real-world data set for geophysical forecasting tasks.

\section{Conclusion and future work}

We introduced a scalable neural architecture for the automated development of proper-orthogonal-decomposition-based long-short-term memory networks (POD-LSTM) to forecast the NOAA Optimum Interpolation Sea-Surface Temperature data set. We implemented aging evolution (AE),  an asynchronous evolutionary algorithm within DeepHyper, an open-source automated machine learning package, to significantly improve its scalability on Theta, a leadership-class HPC system at Argonne's Leadership Computing Facility. We compared AE with the two search methods already within DeepHyper, a distributed reinforcement learning and random search method, and showed that AE outperforms the distributed reinforcement learning method with respect to node utilization and \revised{scalability}. In addition, AE achieves architectures with better accuracy in shorter wall-clock time and matches the node utilization \revised{of the completely asynchronous random search}. We compared the best architecture obtained from AE that was retrained for a longer number of epochs with manually designed POD-LSTM variants and linear and nonlinear forecasting methods. We showed that the automatically designed architecture outperformed all of the baselines with respect to the accuracy on the test data, obtaining a $R^2$ value of $0.876$. The automated data-driven POD-LSTM method that we developed has the potential to provide fast emulation of geophysical phenomena and can be leveraged within ensemble forecast problems as well as for real-time data assimilation tasks. Since the POD-LSTM method relies on the interpretation of flow fields solely from the perspective of data, they may be used for forecasting in multiple applications involving spatiotemporally varying behavior. However, a key point that needs domain-specific intuition is the underlying knowledge of spatial and temporal scales that determines the sampling resolution for snapshots and the dimension of the reduced basis constructed through the POD compression. For example, basic POD-based methods (without task-specific augmentation) fail to capture shocks or contact discontinuities, which are common in engineering applications \cite{taira2020modal}.

Keeping geophysical emulators as our focus, our future work will seek to overcome the limitations of the POD by hybridizing compression and time evolution in one end-to-end architecture search. In addition, we aim to deploy such emulation discovery strategies for larger and more finely resolved data sets for precipitation and temperature forecasting on shorter spatial and temporal scales. The results from this current investigation also are promising for continued investigation into AutoML-enhanced data-driven surrogate discovery for scientific machine learning.

\section*{Acknowledgments}
The authors acknowledge the valuable discussions with Dr. Rao Kotamarthi (Argonne National Laboratory) and Ashesh Chattopadhyay (Rice University) in the preparation of this manuscript. This material is based upon work supported by the U.S. Department of Energy (DOE), Office of Science, Office of Advanced Scientific Computing Research, under Contract DE-AC02-06CH11357. This research was funded in part and used resources of the Argonne Leadership Computing Facility, which is a DOE Office of Science User Facility supported under Contract DE-AC02-06CH11357. RM acknowledges support from the Margaret Butler Fellowship at the Argonne Leadership Computing Facility. This paper describes objective technical results and analysis. Any subjective views or opinions that might be expressed in the paper do not necessarily represent the views of the U.S. DOE or the United States Government. Declaration of Interests - None.

\bibliographystyle{plain}
\bibliography{references}

\begin{center}
    \framebox{\parbox{3in}{
    The submitted manuscript has been created by UChicago Argonne, LLC, Operator of Argonne National Laboratory (``Argonne''). Argonne, a U.S. Department of Energy Office of Science laboratory, is operated under Contract No. DE-AC02-06CH11357. The U.S. Government retains for itself, and others acting on its behalf, a paid-up nonexclusive, irrevocable worldwide license in said article to reproduce, prepare derivative works, distribute copies to the public, and perform publicly and display publicly, by or on behalf of the Government. The Department of Energy will provide public access to these results of federally sponsored research in accordance with the DOE Public Access Plan. \url{http://energy.gov/downloads/doe-public-access-plan}}}
    \normalsize
\end{center}

\end{document}